\documentclass{svproc}
\usepackage{url}

\usepackage{graphicx}
\usepackage{slashed}
\usepackage{multirow}
\usepackage{amsmath}
\usepackage{calligra}
\usepackage[dvipsnames]{xcolor}
\usepackage{calrsfs}
\DeclareMathAlphabet{\mathcalligra}{T1}{calligra}{m}{n}
\DeclareMathAlphabet{\pazocal}{OMS}{zplm}{m}{n}

\def\S{\pazocal{S}}
\def\la{\langle}
\def\ra{\rangle}
\def\to{\rightarrow}

\def\rad{\gamma}

\def\T{\pazocal{T}} 
\def\Lag{\pazocal{L}}
\def\re{t}
\def\im{a}
\def\mre{m_\re}
\def\mim{m_\im}
\def\Ga{\Gamma}
\def\Gre{\Ga_\re}
\def\Gim{\Ga_\im}

\def\dm{\text{\tiny{DM}}}
\def\a{\alpha}
\def\b{\beta}

\def\m{\mu}
\def\n{\nu}

\def\rh{\text{\tiny{RH}}}
\def\max{\text{\tiny{MAX}}}

\def\M{\pazocal{M}}

\definecolor{darkpastelgreen}{rgb}{0.01, 0.75, 0.24}

\begin{document}
\mainmatter              
\title{The moduli portal to dark matter particles}
\titlerunning{The moduli portal to dark matter particles}  
%
\author{Ma\'ira Dutra$^{[0000-0003-4851-242X]}$}
\authorrunning{Ma\'ira Dutra} 
%
\tocauthor{Ma\'ira Dutra}
\institute{Carleton University, Ottawa ON K1S 5B6, Canada,\\
\email{mdutra@physics.carleton.ca},
}

\maketitle              

\begin{abstract}
The out-of-equilibrium production of dark matter (DM)  from standard model (SM) species in the early universe (freeze-in mechanism) is expected in many scenarios in which very heavy beyond the SM fields act as mediators. In this conference, I have talked about the freeze-in of scalar, fermionic and vector DM though the exchange of moduli fields \cite{chowdhury_moduli_2019}, which are in the low-energy spectrum of many extra-dimensions and string theory frameworks. We have shown that the high temperature dependencies of the production rate densities in this model, as well as the possibility of having moduli masses at the post-inflationary reheating scale, make it crucial to consider the contribution of the freeze-in prior the start of the standard radiation era for a correct prediction of the DM relic density. 

\keywords{Dark matter; Moduli fields; Freeze-in production}
\end{abstract}
\section{Introduction}

The close relationship between the couplings dark matter (DM) particles might have to particles belonging to the Standard Model of Particle Physics (SM) and their evolution through the early universe makes the DM puzzle an open problem in the interface between particle physics and cosmology.

Direct detection searches seek to detect nuclear or electronic recoils from DM scatterings, and the current status of no positive signals but more and more sensitive detectors mean that the SM-DM coupling need to be weaker and weaker. However, very weak couplings might imply that DM and SM particles were never in thermal equilibrium in the early universe. The out-of-equilibrium production of DM from SM species in the early universe, which is the so-called \textit{freeze-in mechanism} \cite{chung_production_1998,hall_freeze-production_2010,bernal_dawn_2017}, is expected in many scenarios in which very heavy beyond the SM (BSM) fields act as mediators. If the masses of these heavy mediators are close to the scale of the post-inflationary reheating, it is important to take into account the freeze-in during a reheating period in which entropy is being injected into the thermal bath. Here we shed light on this matter, presenting handy formulae that can be useful for any scenario.

Moduli fields are scalars which would be present in the effective limit of many string theory frameworks. Since they would need to be very feebly coupled to SM fields, it is interesting to investigate whether their feeble interactions with dark and visible matter would be
enough to produce dark matter via freeze-in. In this conference, I have presented a recent study on the moduli portal to dark matter \cite{chowdhury_moduli_2019}.

%
\section{The model}

We have considered a complex modulus field, $\T = t + i a$, whose presence at temperatures below some cut-off scale $\Lambda$ would appear as corrections to the free Lagrangians\footnote{We here consider corrections up to the first order in the cut-off scale $\Lambda$.}. We consider BSM scalar, fermion and vector fields as feebly interacting massive particle (FIMP) candidates, which are DM candidates produced via freeze-in. Our effective Lagrangian connecting the modulus field to dark and standard scalars, fermions and vectors, here generically denoted by $\Phi, \Psi$ and $X_\mu$, reads
\begin{equation}
\Lag_{eff} \supset 
\begin{cases}
\left(1 + \frac{\a_i}{\Lambda}t\right) (|D_\mu \Phi|^2 - \mu_\Phi^2)     & \text{(scalars)} \\ 
\frac{1}{2}\left(1 + \frac{\a_{L,R}^i}{\Lambda}t + i \frac{\beta_{L,R}^i}{\Lambda}a\right) \bar{\Psi}_{L,R} i \slashed{D} \Psi_{L,R} 
    & \text{(fermions)} \\
-\frac{1}{4} \left(1+ \frac{\a_i}{\Lambda}t\right) X_{\mu \nu} X^{\mu \nu} - \frac{\beta_i}{\Lambda}a \, X_{\mu \nu} \tilde X^{\mu \nu} 
    &\text{(vectors)} \\ 
\end{cases}
\end{equation}

Couplings to the real and imaginary components of the modulus are denoted respectively by $\a$ and $\b$, and the tensors $X_{\m \n}$ and $\tilde X_{\m \n}$ are respectively the field strength and dual field strength of $X_\m$.

In order to avoid imaginary contributions to the mass and kinetic terms of the scalars, we assume that only the real component of the modulus interact with the SM Higgs and the scalar FIMP candidate. In the case of the scalar FIMP, though, we do not assume that the modulus change the mass term. The interactions with fermions are in principle chiral and a chirality flip would give us explicit dependence on the fermion mass in the amplitudes. For this reason, SM fermions cannot produce the FIMPs above EWSB, since they are massless. For the interactions of moduli with vectors, we have a Higgs-like operator for the real component and a Peccei-Queen operator for the axial component.

\begin{figure}[t!]
\centering
\includegraphics[scale=0.4]{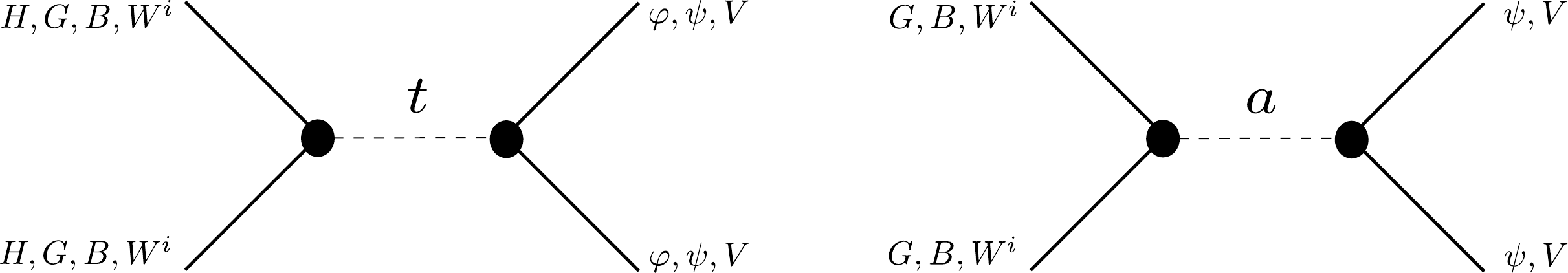}
\caption{Schematic Feynman diagrams leading to the freeze-in production of our FIMP candidates, scalar ($\phi$), fermionic ($\psi$) and vector ($V_\mu$) fields, out of annihilations of all the SM bosons ($H, G_\mu^a,B_\mu, W_\mu^i$) via exchange of the real ($t$) and axial ($a$) components of a modulus field.}
\label{fig:diagrams}
\end{figure}

As we are going to emphasize in the next section, the squared amplitudes of the freeze-in processes give us valuable information about the freeze-in temperature. In our case, the FIMP candidates are produced from $s$-channel annihilations of Higgs bosons and SM gauge bosons, having $t$ and $a$ as mediators, as depicted in Fig. \ref{fig:diagrams}. The squared amplitudes are given by:
\begin{equation}\label{M0}
|\M|^2_0 = \frac{\a_\dm^2 \lambda (s,\a_i)}{\Lambda^4} \frac{ s^4 \Big( 1 - \frac{2m_\dm^2}{s} \Big)^2}{(s-\mre^2)^2+\mre^2 \Gre^2}
\end{equation}

\begin{equation}\begin{split}\label{M12}
|\M|^2_{1/2} = & \frac{\a_\dm^2 \lambda (s,\a_i)}{\Lambda^4} \frac{ m^{2}_\dm s^3 \Big( 1 - \frac{4m_\dm^2}{s} \Big)}{(s-\mre^2)^2+\mre^2 \Gre^2}
+ \frac{\beta_\dm^2 \lambda (s,\b_i)}{\Lambda^4} \frac{ m_\dm^2 s^3}{(s-\mim^2)^2+\mim^2 \Gim^2}
\end{split}\end{equation}

\begin{equation}\begin{split}\label{M1}
|\M|^2_{1} = & \frac{\a_\dm^2 \lambda (s,\a_i)}{\Lambda^4}\frac{ s^4 \Big( 1 - \frac{4 m_\dm^2}{s} + \frac{6m_\dm^4}{s^2} \Big)}{(s-\mre^2)^2+\mre^2 \Gre^2} 
+ \frac{\beta_\dm^2 \lambda (s,\b_i)}{\Lambda^4} \frac{ s^4 \Big( 1 - \frac{4 m_\dm^2}{s} \Big)}{(s-\mim^2)^2+\mim^2 \Gim^2} \,. 
\end{split}\end{equation}

Above, $\lambda (s,\a_i), \lambda (s,\b_i)$ are the sums over Higgs and SM gauge bosons contributions, which can be a function of the Mandelstam variable $s$. For further details, see \cite{dutra_origins_2019}.

\section{Results}
\subsection{Production rate densities and evolution of FIMP relic density}

The freeze-in temperature is determined by the interactions between FIMPs and the SM species involved. In particular, the temperature-dependence of the production rate densities tell us if the freeze-in happens at the lowest or highest scale of a given cosmological period. This is what I want to emphasize in this section.

The rate at which the number of DM particles change in a comoving volume $a^3$, with $a$ the scale factor, is given by the Boltzmann fluid equation $\dot N_\dm = R_\dm (t) a^3$, with $N_\dm = n_\dm a^3$ the total number of DM particles and $R_\dm (t)$ the time/temperature-dependent interaction rate density, which in the case of the freeze-in only account for production and not for loss of FIMPs. On the other hand, the Hubble rate $H(t)$ determines how the scale factor varies with time, $H(t) = \dot a/a$, and since this quantity is proportional to the total energy density of the universe, different species dominating the expansion lead to different final total number of DM particles, as well as different time-temperature relations.

For a $12 \to 34$ process, the production rate density of species $3$, in the limiting case where species $1$ and $2$ have Maxwell-Boltzmann distributions, is given by \cite{dutra_origins_2019}
\begin{equation}\begin{split}
R_3^{12\to 34} \equiv n_1^{eq} n_2^{eq} \la \sigma v \ra = \frac{\S_{12}\S_{34}}{32(2\pi)^6}\int & ds \frac{\sqrt{\lambda(s,m_3^2,m_4^2)}}{s} \int d\Omega_{13} |\M|^2 \\
& \times
\frac{2T}{\sqrt{s}} \sqrt{\lambda(s,m_1^2,m_2^2)} K_1\left(\frac{\sqrt{s}}{T}\right)\,, 
\end{split}\end{equation}

\begin{figure}[t]
\centering
\includegraphics[scale=0.3]{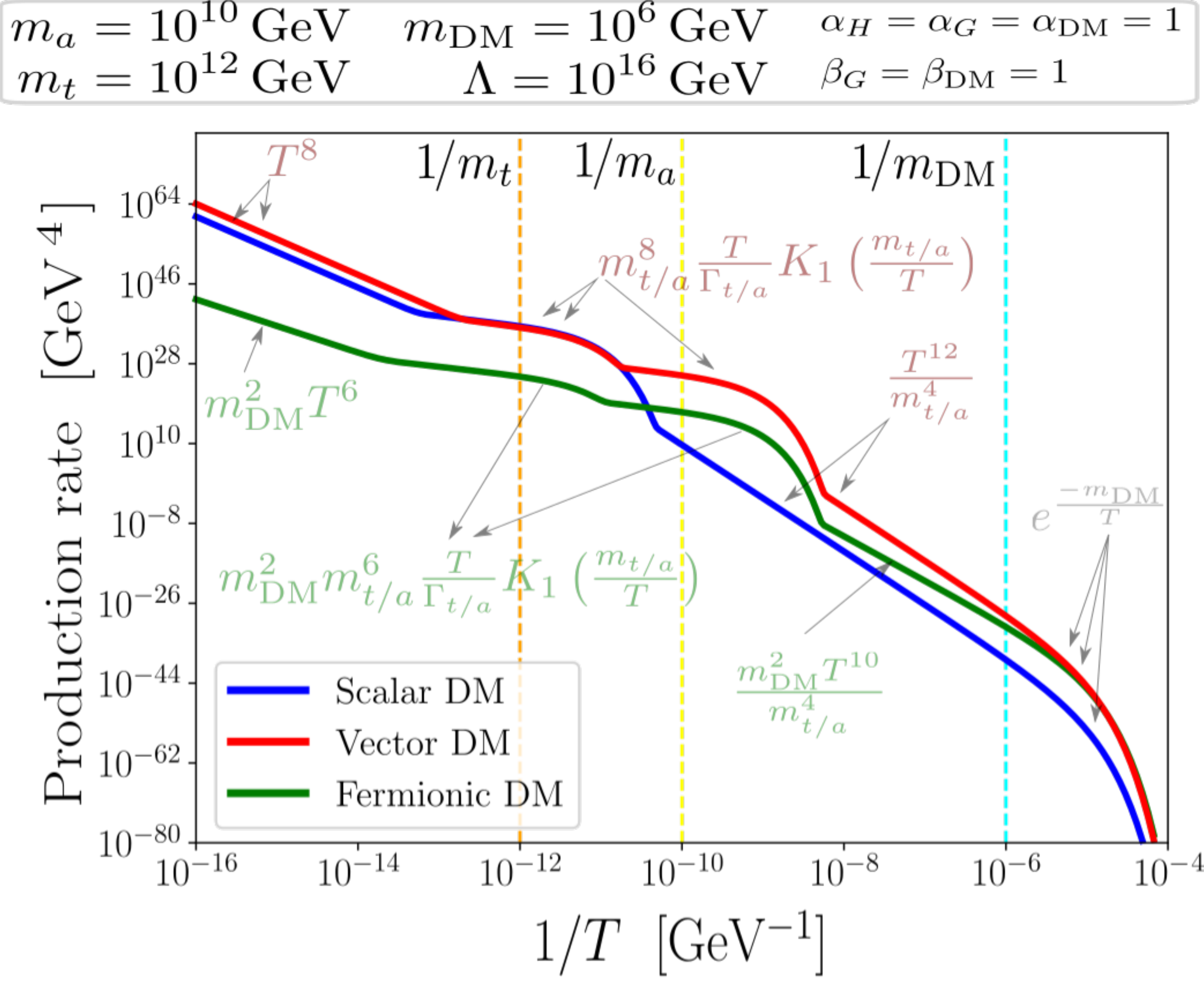}
\caption{Production rate densities of our FIMP candidates as functions of the inverse of the SM bath temperature.}
\label{fig:rates}
\end{figure}

In Fig.\ref{fig:rates}, I show the production rate densities of a scalar, fermionic and vector FIMPs (blue, green and red curves respectively), as functions of the inverse of temperature. Generic features of the production rates are that, the higher the temperature, the more DM is produced. Also, notice that the rates start to vanish when the temperature
of the thermal bath becomes smaller than the DM mass, which is the famous Boltzmann suppression. We can recognize the presence of the poles of both components of the modulus, when the temperature of the thermal bath equals their masses. We can also notice the weaker temperature dependence of the production rate of a fermionic FIMP, due to the chirality flip.
For completeness, I point out in Fig. \ref{fig:rates} the analytic approximations of the production rates in the limiting cases where the mediators are much lighter, of the same order and much heavier than the temperature of the SM thermal bath. 

As previously stated, the relic density of DM depends on which species dominates the cosmic expansion. While it is usual to assume that DM production happened during a radiation-dominated era, in which $H(T) \propto T^2$, this might not be the case in general. In inflationary theories, the universe had undergone a period of entropy production called \textit{reheating} in which it cools down slower and $H(T) \propto T^4$. Such a period would happen from a moment when the temperature of the SM bath reaches a maximal value $T_\max$ up to the moment in which there is no more entropy production, the so-defined reheat temperature $T_\rh$. We do not know the scale of $T_\rh$, which could be as low as $4 \times 10^{-3}$ GeV \cite{hannestad_what_2004} and as high as $7 \times 10^{15}$ GeV \cite{rehagen_low_2015}. A general study of freeze-in through heavy portals should therefore take into account the possibility that the masses of mediators are at the reheating scale. So, as long as we have a thermal bath of SM radiation, it starts producing DM. In this context, the relic density of dark matter today receives a contribution from the reheating period and from the radiation era \cite{dutra_origins_2019}: 
\begin{equation}\begin{split}\label{relicsplit}
\Omega_\dm^0 h^2 
= \frac{m_\dm}{2.16 \times 10^{-28}} \Big( & \int_{T_0}^{T_\rh} dT \frac{g_s^*}{g_s \sqrt{g_e}} \frac{R_\dm (T)}{T^6} + \\
&+ 1.6\,c\,B_\rad  \,g_\rh^{-3/2}\, T_\rh^7 \int_{T_\rh}^{T_\max} dT g_e^* \frac{R_\dm (T)}{T^{13}} \Big) \, .
\end{split}\end{equation}

It is easy to see from the equation above that if $R_\dm \propto T^n$ for $n<5$, the production during radiation era is infrared (happens at the lightest scale available) and if $n>5$, it is ultraviolet (happens at the highest scale available, $T > T_\rh$). The same is true for the production during reheating, where the power of temperature is $n = 12$. 

From the temperature dependence of the production rates pointed out in Fig.\ref{fig:rates}, we can understand that the freeze-in of all the FIMP candidates happen at the highest scale of radiation era, the reheating temperature ($n>5$), except for the case where the moduli are produced on-shell, which can make the relic density raise again after levelling-off. Neglecting the production during reheating in such a model lead to an underestimation in the relic density of many orders of magnitude, as it was explicitly shown in the case of an on-shell exchange of spin-2 fields \cite{bernal_spin-2_2018}.

\begin{figure}[t!]
\centering
\includegraphics[scale=0.3]{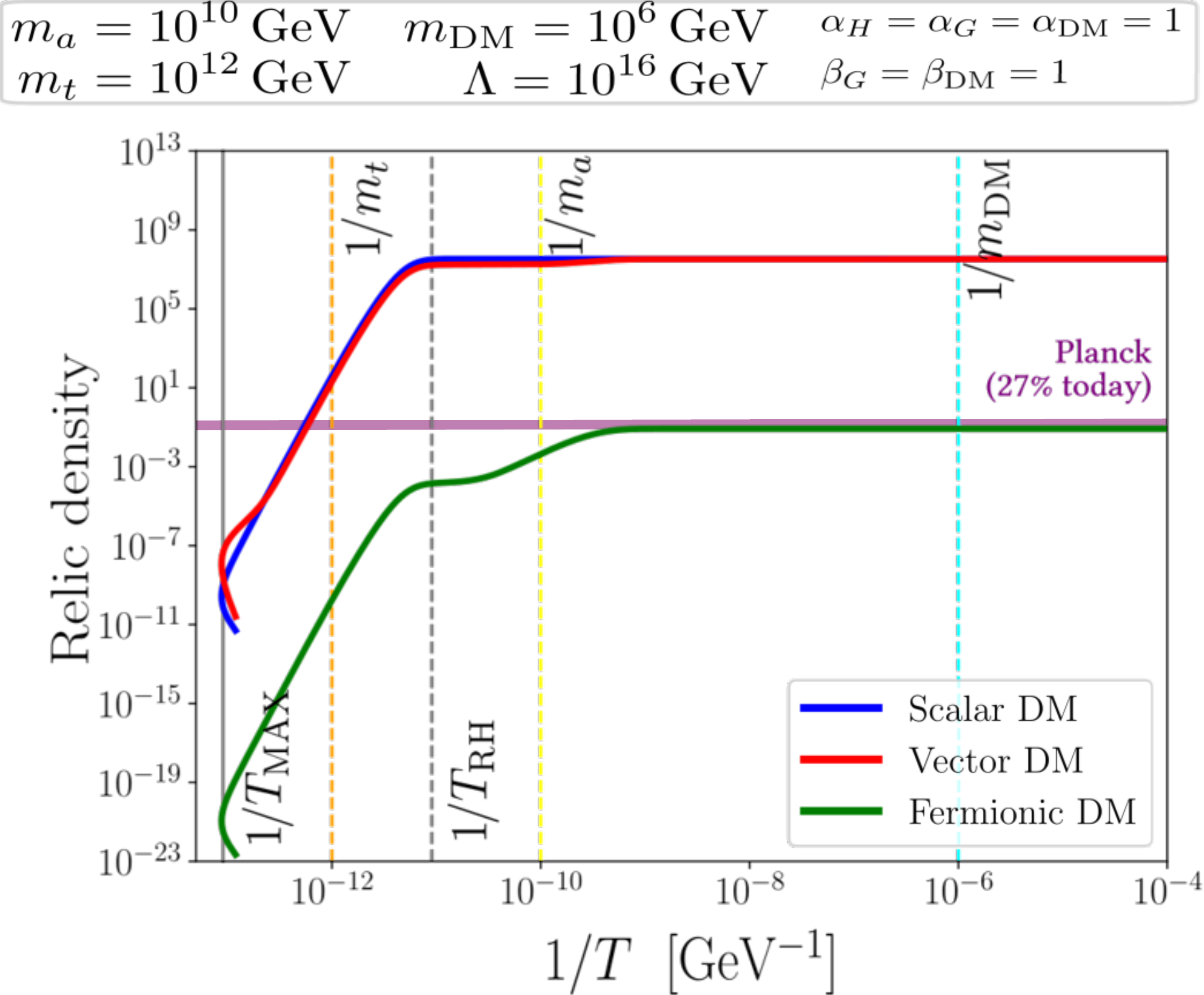}
\caption{Evolution of the relic density of our FIMP candidates.}
\label{fig:relic}
\end{figure}

In Fig. \ref{fig:relic}, we see the resulting evolution of the relic density, for the same set of free parameters of Fig. \ref{fig:rates}. These are the solutions of the coupled set of Boltzmann fluid equations for the evolution of DM, SM radiation and inflaton (driving the reheat period). We have fixed $T_\rh = 10^{11}$ GeV and $T_\max = 10^{13}$ GeV. In the presence of an on-shell production of a
mediator, the production of DM is enhanced and that is why we see the relic density of vector and fermionic DM getting enhanced close to the pole of the axial
modulus. Of course, the final relic density needs to agree to the Planck results, as I am going to show in the next section.

\subsection{Agreement with Planck results}

\begin{figure}[t!]
\centering
\includegraphics[scale=0.35]{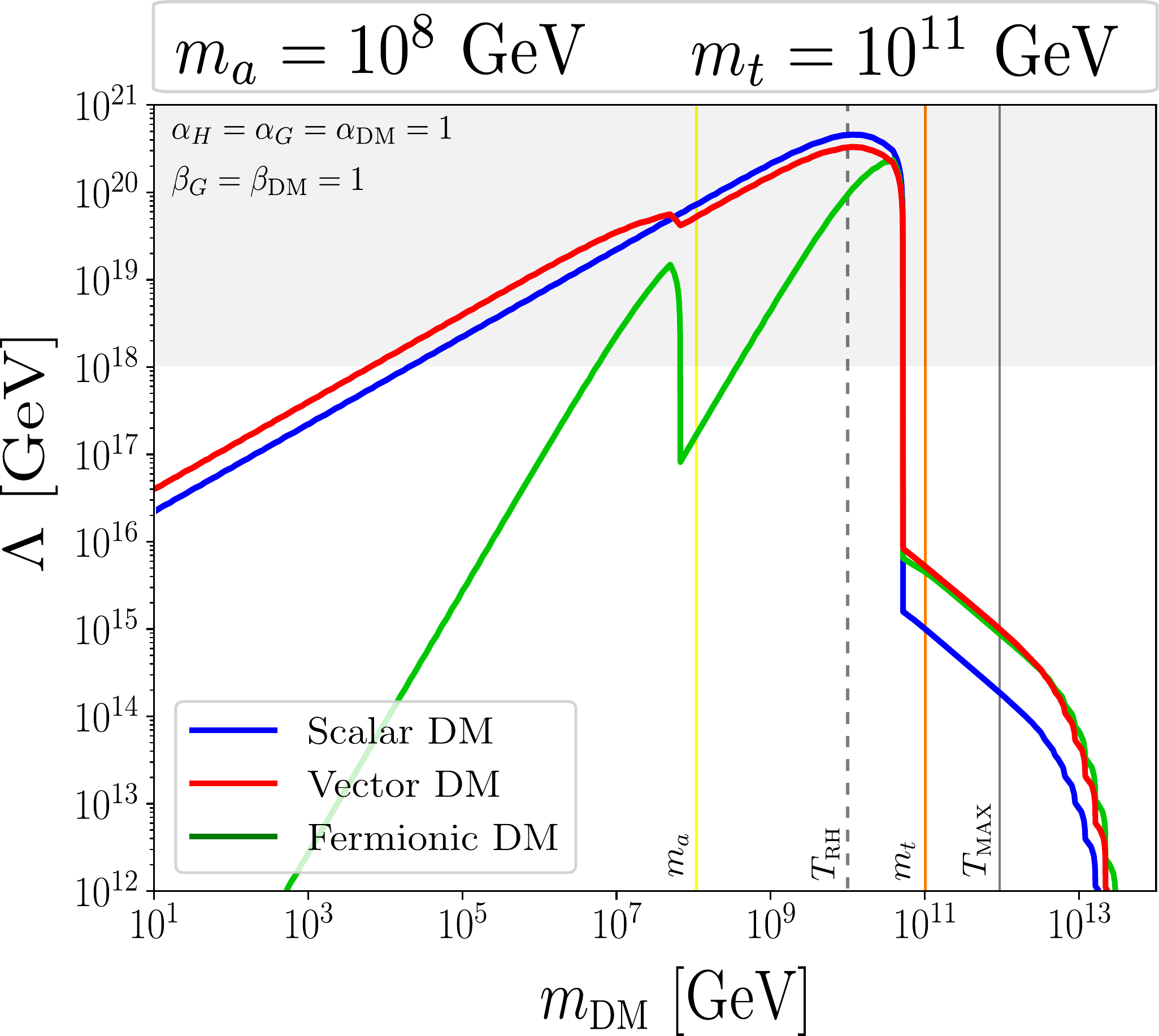}

\vspace{.3cm}
\includegraphics[scale=0.35]{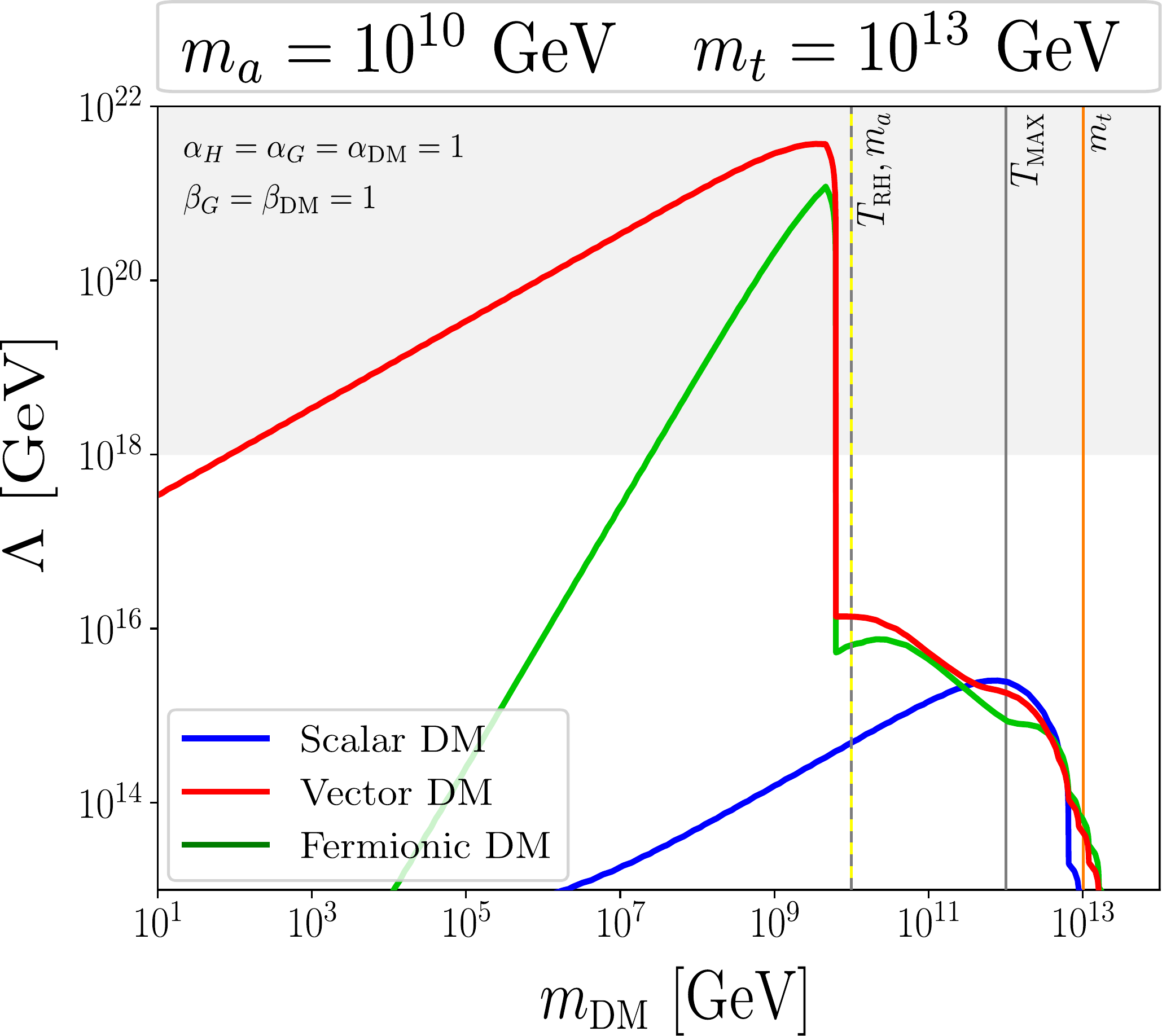}
\caption{Contours of good relic density in our free parameter space.}
\label{fig:contours}
\end{figure}

We can now see the values of the new physics scale $\Lambda$ and FIMP mass providing a good relic density of DM today, as inferred by the Planck satellite \cite{aghanim_planck_nodate}. In Fig. \ref{fig:contours}, the reheating and maximal temperatures are set to $10^{10}$ and $10^{12}$ GeV.
Since the relic density increases with DM mass, we need smaller overall couplings in order to not overproduce the FIMPs and therefore $\Lambda$ needs to be raised for a given relic density value. Also, due to the exponential Boltzmann suppression, the thermal
bath cannot produce dark matter much before the time of maximal temperature and $\Lambda$ is 
sharply lowered as to compensate the suppressed rates.
In the upper panel, the pole of the axial modulus is reached inside the radiation era. Since
it enhances the relic densities of vector and fermionic DM candidates, we need higher
values of $\Lambda$. Notice that the curves for the fermionic DM depend strongly on the DM
mass, due to the chirality flip.
In the lower panel, the exchange of a heavier real modulus suppresses more the relic densities, so that we can have lower values of $\Lambda$, but still at intermediate scales.

\section{Concluding remarks}

In this conference, I have discussed the case in which heavy moduli fields exchange between visible and dark matter are the underlying physics of the feeble couplings necessary for the freeze-in to happen. Such fields appear in many structural extensions of the SM, and our results are expected to be embedded in more
realistic realizations. We have seen that if the temperature dependencies of the production rates of FIMPs is strong enough, which can be achieved in effective models with derivative couplings, FIMPs would have already been produced at the start of the radiation era. As an interesting outcome, in a wide range of our parameter space a good relic density is “naturally”
achieved for scalar, fermionic and vector FIMP candidates with the moduli masses at intermediate
scales, and for reasonable scales of new physics.

\subsubsection{Acknowledgments.} I want to thank the co-authors of the work presented in this conference, Debtosh Chowdhury, Emilian Dudas and Yann Mambrini. I also acknowledge the support from the Brazilian PhD program “Ciências sem Fronteiras”-CNPQ Process No. 202055/2015-9 during the development of this work and the current support of the Arthur B. McDonald Canadian Astroparticle Physics Research Institute. 

%

\end{document}